\documentclass[conference]{IEEEtran}
\usepackage{cite}

\ifCLASSINFOpdf
\usepackage[pdftex]{graphicx}
\else
\fi
\usepackage[cmex10]{amsmath}
\usepackage{algorithmic}

\usepackage{array}

\usepackage[caption=false,font=footnotesize]{subfig}
\usepackage{subfig}

\usepackage{fixltx2e}

\usepackage{stfloats}
\usepackage{url}

\begin{document}
%
\title{\LARGE{Evolving Optical Networks for Latency-Sensitive Smart-Grid Communications via Optical Time Slice Switching (OTSS) Technologies}}


%
\author{\IEEEauthorblockN{Zhizhen Zhong\IEEEauthorrefmark{1},
Nan Hua\IEEEauthorrefmark{1},
Zhu Liu\IEEEauthorrefmark{2},
Wenjing Li\IEEEauthorrefmark{2},
Yanhe Li\IEEEauthorrefmark{1} and
Xiaoping Zheng\IEEEauthorrefmark{1}
}
\IEEEauthorblockA{\IEEEauthorrefmark{1}Tsinghua National Laboratory for Information Science and Technology (TNList), \\
Department of Electronic Engineering, Tsinghua University, Beijing, P. R. China.}
\IEEEauthorblockA{\IEEEauthorrefmark{2}State Grid Information \& Telecommunication Group Co., LTD., Beijing, P. R. China.}
\url{xpzheng@mail.tsinghua.edu.cn}
}

\IEEEspecialpapernotice{(IEEE Photonics Society 1st Place Best Poster Award, on CLEO-PR/OECC/PGC 2017)\\(Camera-ready version for arXiv only)}

\maketitle

\begin{abstract}
In this paper, we proposed a novel OTSS-assisted optical network architecture for smart-grid communication networks, which has unique requirements for low-latency connections. Illustrative results show that, OTSS can provide extremely better performance in latency and blocking probability than conventional flexi-grid optical networks.
\end{abstract}


%
\IEEEpeerreviewmaketitle

\section{Introduction}
Next-generation power grid, known as the smart grid, renovates the conventional paradigm of electricity distribution and management to accomplish intelligent control, improved efficiency and reliability \cite{yan-survey-2013}. New applications, e.g. smart metering systems, wide-area situational awareness systems and distribution automation systems, along with its related intense and diverse traffic flows propose new demands on communication infrastructures \cite{Gungor-trans-2013}. In smart grid communication networks, the Quality of Service (QoS) among power suppliers, remote power infrastructures and customers is very important, for any service degradation (latency, outage) in reliability-related traffic (e.g. sensor data, control signals and monitors' video flows) may compromise stability. Therefore, we should try to lower the service latency as much as possible.

Optical network is regarded as a promising candidate for smart grid communications due to its unique advantages, such as high volume, low latency and jitter. As optical lightpaths usually have much larger capacity than the bandwidth of a specific traffic request, in most cases, fine-grained traffic requests are accommodated via traffic grooming \cite{zhu-jsac-2002}, while lightpaths are provisioned over the physical fiber topology, which compose the multi-layer network architecture. However, conventional multi-layer networks, i.e. IP-over-WDM networks, have strict bandwidth constraints on the optical layer, so that a fine-grained traffic may trigger a new lightpath with redundant capacity, and soon exhaust the available spectrum, then other traffic has to be accommodated by traffic grooming, which adds to service latency due to queueing and multiplex/demultiplex at grooming nodes.

As a major breakthrough in the optical layer, elastic optical networking, or flexi-grid optical networking enables more flexibility in optical spectrum assignment \cite{Gerstel-magazine-2012}. A significant advantage of elastic optical networks is that it can provide fine-granularity lightpaths to avoid the waste of resources, while optical super-channels with ultra-high bandwidth are also available to save transceivers. However, flexi-grid technology still meets obstacles when even finer granularity is needed, for instance, in smart grid where traffic requests are strict in latency and small in bandwidth. Luckily, Optical Time Slice Switching (OTSS) is proposed as a novel all-optical switching technology that exploits temporal domain to provide more transparent connections \cite{otss-acp} \cite{patent}. Commercial PLZT fast switches \cite{PLZT} make OTSS a practical technology, and experimental demonstration has been conducted in \cite{ofc2016}. 

\section{Traffic Model in Smart-Grid Communication Networks (SGCN)}
In Smart-Grid Communication Networks (SGCN), diverse kinds of traffic requests are generated among massive entities ubiquitously. Briefly, there are three paradigms \cite{bell-journal}.

\subsubsection{Hub-and-Spoke (HS)}
Smart Metering is a typical hub-and-spoke traffic. The grid control center collects operating messages (e.g., voltage, frequency) and billing information (e.g., real-time pricing (RTP), time of use (TOU) pricing, and critical peak pricing (CPP)) at certain intervals. This kind of traffic is distributed between all substations and the control center.

\subsubsection{Peer-to-Peer (P2P)}
Teleprotection traffic is generated in a peer-to-peer manner between substations along power-transmission lines. A fault at one substation can be remotely detected within extremely short time window via teleprotection traffic at adjoining substations, which is crucial to the stability of grid infrastructure. IEEE 1646 \cite{IEEE1646} and International Electrotechnical Commission (IEC) 61850 require communication latency of teleprotection communications to be as little as 1/2 of a cycle (about 8 ms and 10ms, for 60Hz and 50Hz AC frequencies, respectively).

\subsubsection{Random Distributed}
Traffic requests are distributed uniformly and independently among all node pairs. General purposes for communication in smart-grid communication networks, e.g. file transfer and video streaming, can be regarded as this kind of communication paradigm.

Finally, Table I summarizes the main characteristics of smart-grid traffic requirements in SGCN.

\begin{table}[!t]
\caption{Typical traffic requirements in SGCN \cite{IEEE1646}\cite{Deshpande2011}}
\label{table_example}
\centering
\begin{tabular}{|m{0.9in}<{\centering}|c|c|c|}
\hline
Application & Paradigm & Bandwidth & Latency\\
\hline
Teleprotection & P2P & $\sim$ 500 Kb/s & 8-10 ms\\
\hline
Load Shedding for Underfrequency & P2P \& HS & $\sim$ 500 Kb/s & 10 ms\\
\hline
SCADA & P2P \& HS & $\sim$ 800 Kb/s & 100-200 ms\\
\hline
Smart Metering & HS & $\sim$ 500 Kb/s & 250-1000 ms\\
\hline
File Transfer & Random &  200-1000 Mb/s &  $\geq$ 1000 ms\\
\hline
\end{tabular}
\end{table}

\begin{figure*}[t]
\centering
\subfloat[Illustration of Optical Time Slice Switching (OTSS).]{
\hspace{-3em}
\includegraphics[width=4.3in]{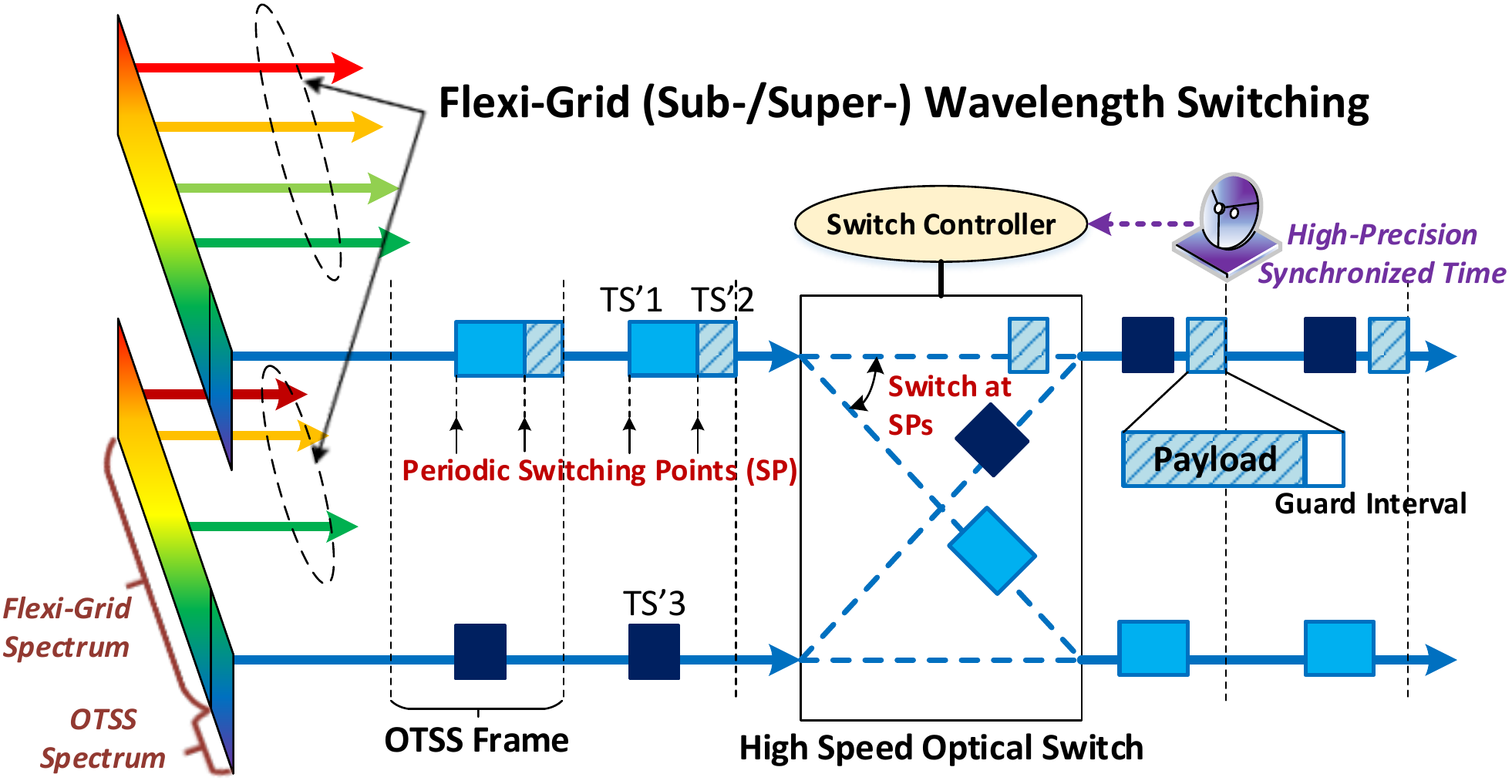}}
\hspace{2em}
\subfloat[OTSS-assisted BV-ROADM.]{
\includegraphics[width=2in]{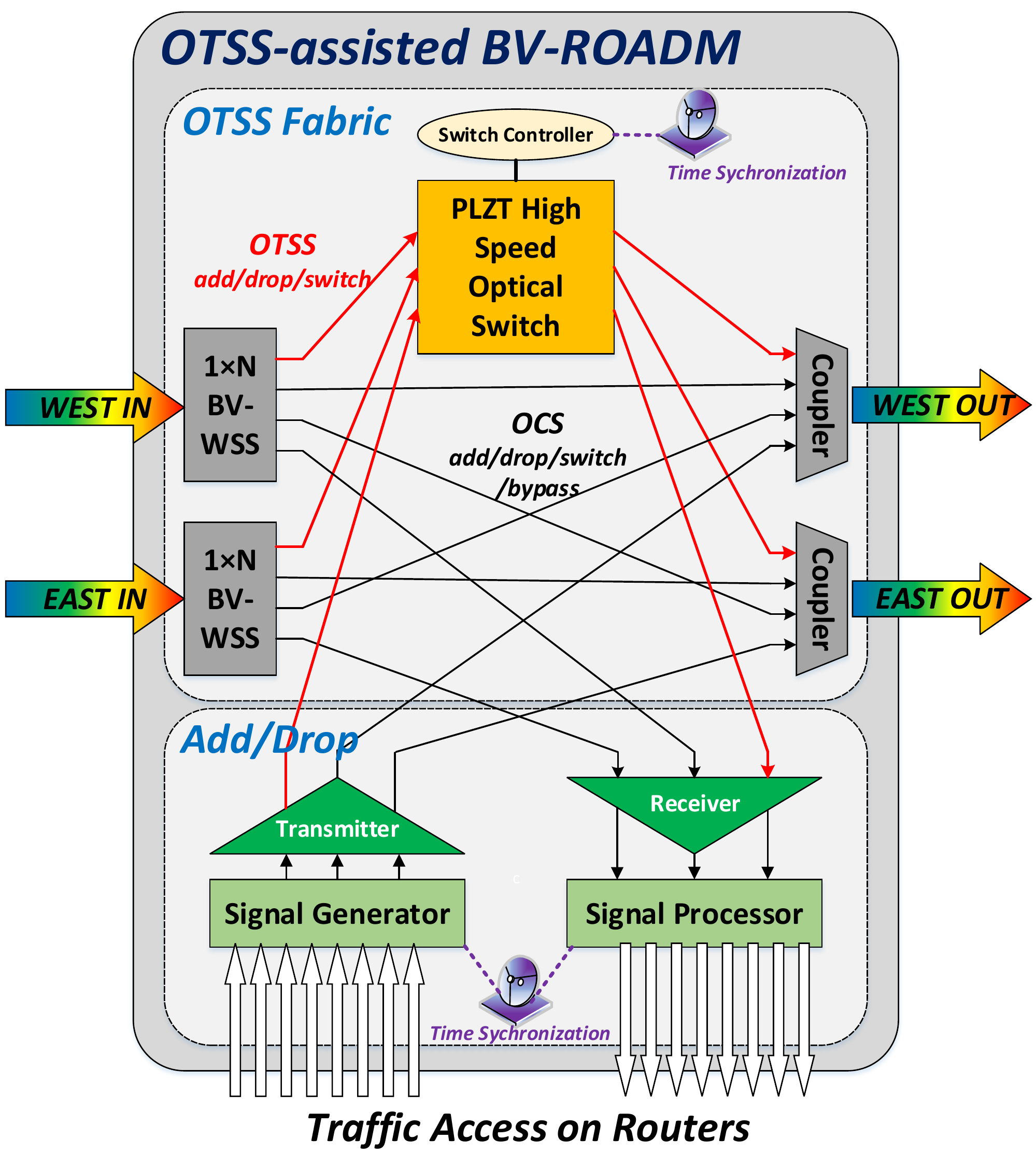}}
\hspace{0.3em}
\caption{OTSS-assisted low-latency flexi-grid optical network architecture.}
\label{fig:subfig} 
\end{figure*}

\section{OTSS-Assisted Low-Latency Optical Network Architecture for Smart Grid}
Due to the latency-sensitive characteristic of smart grid communication networks, a novel all-optical fine-grained switching paradigm, named Optical Time Slice Switching (OTSS), is introduced.

Fig. 1(a) illustrates the principle of OTSS, and detailed explanation can be found in \cite{otss-acp}\cite{patent}. Note that high-precision time synchronization is the key enabling technology for OTSS. With the help of synchronized time at every switching node, we can achieve accurate time-slice allocation within OTSS frame to avoid collision. A good news is that nowadays power grid is already equipped with time-synchronization systems, driven by Global Positioning System (GPS) or IEEE 1588 Precision Time Protocol (PTP), which can satisfy the demand of OTSS with precision of 10 ns synchronization. Note that propagation delay should be considered in time slice allocation at different nodes, and this problem has been solved in \cite{otss-acp} by shifting associated time slice in different segments of a path. The allocation of time slots should also obey the time-slot-continuity constraint, which ensures that an OTSS connection traversing multiple links does not change its relative place within a OTSS frame, when considering propagation delay.

Fig. 1(b) shows the node architecture for OTSS-assisted flexi-grid optical networks. We can evolve current flexi-grid networks by reserving a portion of spectrum for OTSS and adding fast optical switches, while other parts of spectrum still obey conventional wavelength/spectrum-slot routing by wavelength-selective switch. The proposed OTSS-assisted architecture can provide fine-grained connections multiplexed in temporal domain for massive low-latency communications.

\begin{figure}[!b]
\centering
\includegraphics[width=3.4in]{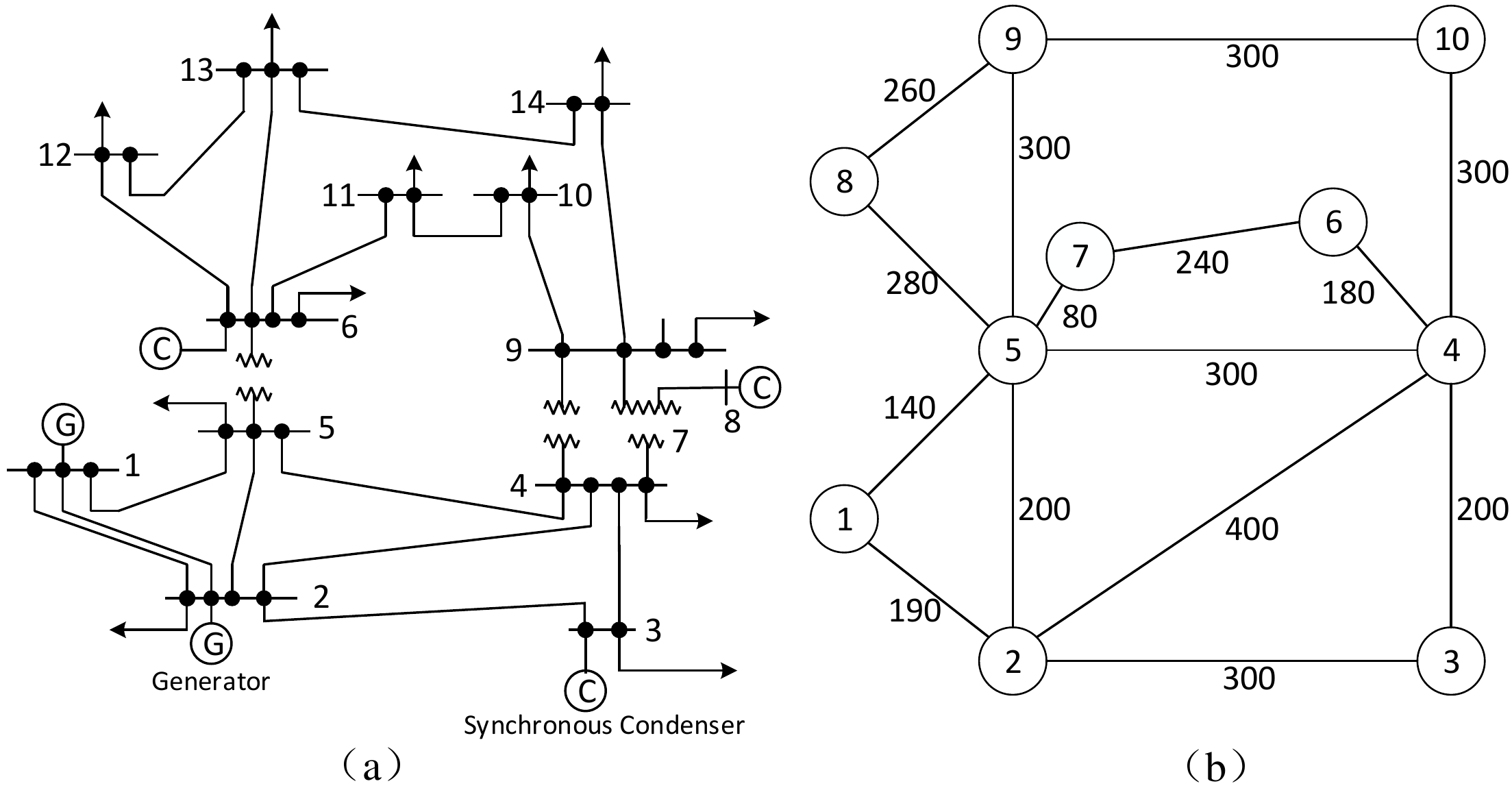}
\caption{IEEE 14 bus system, in which (a) is the power system, (b) is the communication infrastructure, with distance  (kilometer) marked on the graph.}
\label{fig_sim}
\end{figure}

\section{Illustrative Numerical Results}
\subsection{Simulation Setup}
In this study, we consider the well-known IEEE 14-bus system as the smart-grid topology. In power grids, communication links were created in parallel with the transmission lines, resulting in a communication network with almost the same topology as the power system grid (Fig. 2) \cite{IEEE14}.

As reliability traffic poses the main challenge for smart-grid communication networks, we focus on these traffic (teleprotection and load shedding for underfrequency) with the most strict latency requirements in this study. As we simultaneously consider P2P and HS traffic, requests are then generated between all node pairs uniformly and independently, characterized by Poisson arrivals with negative exponential holding times. We assume that the bandwidth of each requests is 500Kb/s, and the upper bound for latency is 10 ms, according to Table I. 

In optical spectrum, 50GHz is reserved for OTSS to accommodate reliability-related traffic. Spetral efficiency is 1 b/s/Hz. For fairness, we also reserve the same amount of spectrum for flexi-grid networks as benchmark to compare its performance with OTSS. On flexi-grid optical networks, four grid sizes are considered (50 GHz, 25 GHz, 12.5 GHz, 6.25 GHz). On OTSS, we set the OTSS frame to be 1 ms, and the smallest time slice to be 10 $\mu$s \footnote{It has been experimentally demonstrated that PLZT fast optical swich can handle time slice as small as 1 $\mu$s in OTSS systems \cite{ofc2016}.}.

On latency composition, both propagation delay and grooming delay are considered. Generally, optical signal propagates in fibers with the speed of 200 km/ms, which results in propagation delay, and grooming delay is caused by electric processing (queueing, multiplex/demultiplex, etc.). We assume that each grooming node brings 5 ms delay to end-to-end latency for simplicity. Traffic grooming only happens at flexi-grid networks to multiplex fine-grained traffic requests. As OTSS already provide enough transparent connections, we do not allow grooming in OTSS to reduce latency.

\subsection{Dynamic Analysis}

\begin{figure*}[t]
\centering
\subfloat[Maximum latency vs. traffic load.]{
\includegraphics[width=2.25in]{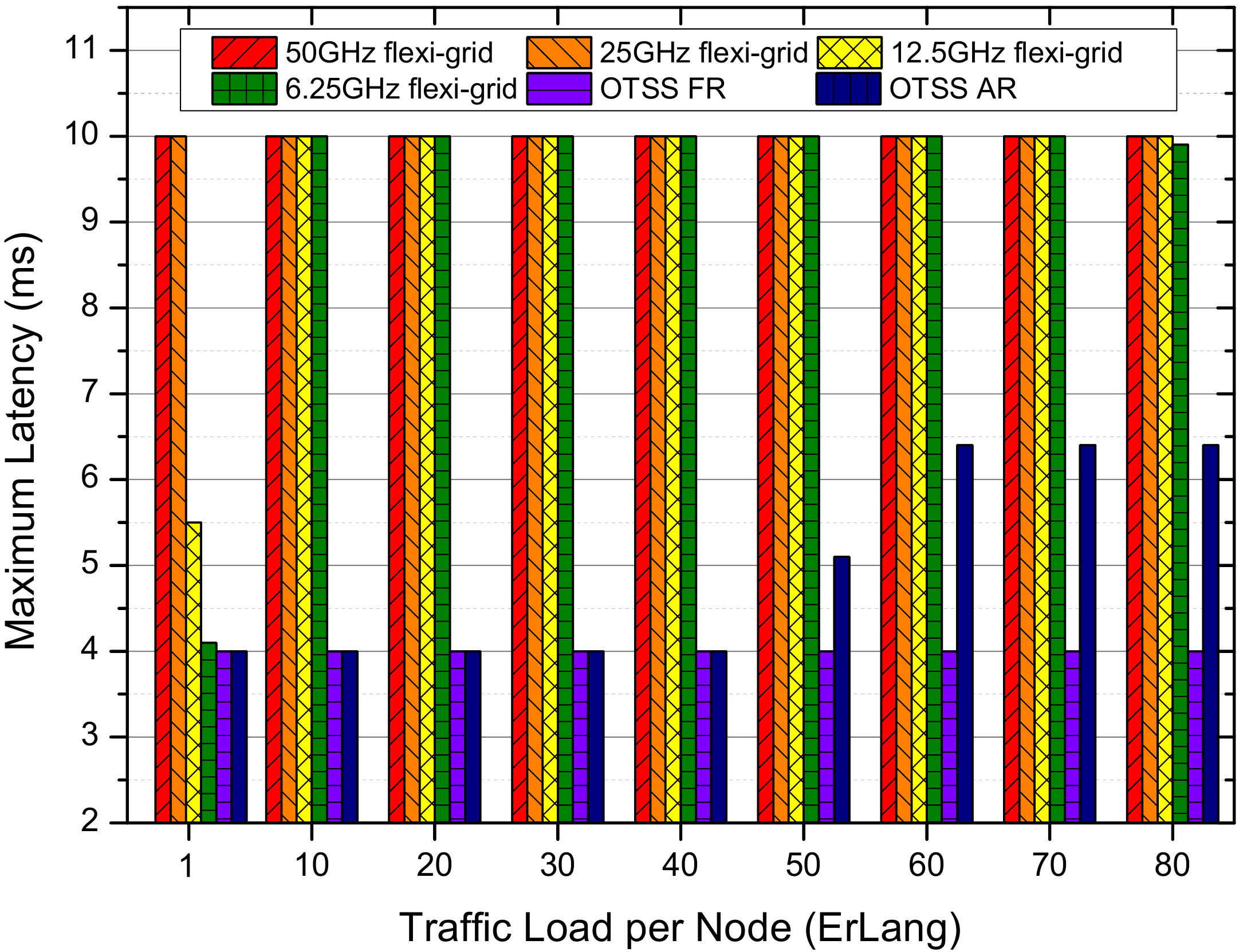}}
\hspace{0.1em}
\subfloat[Average latency vs. traffic load.]{
\includegraphics[width=2.25in]{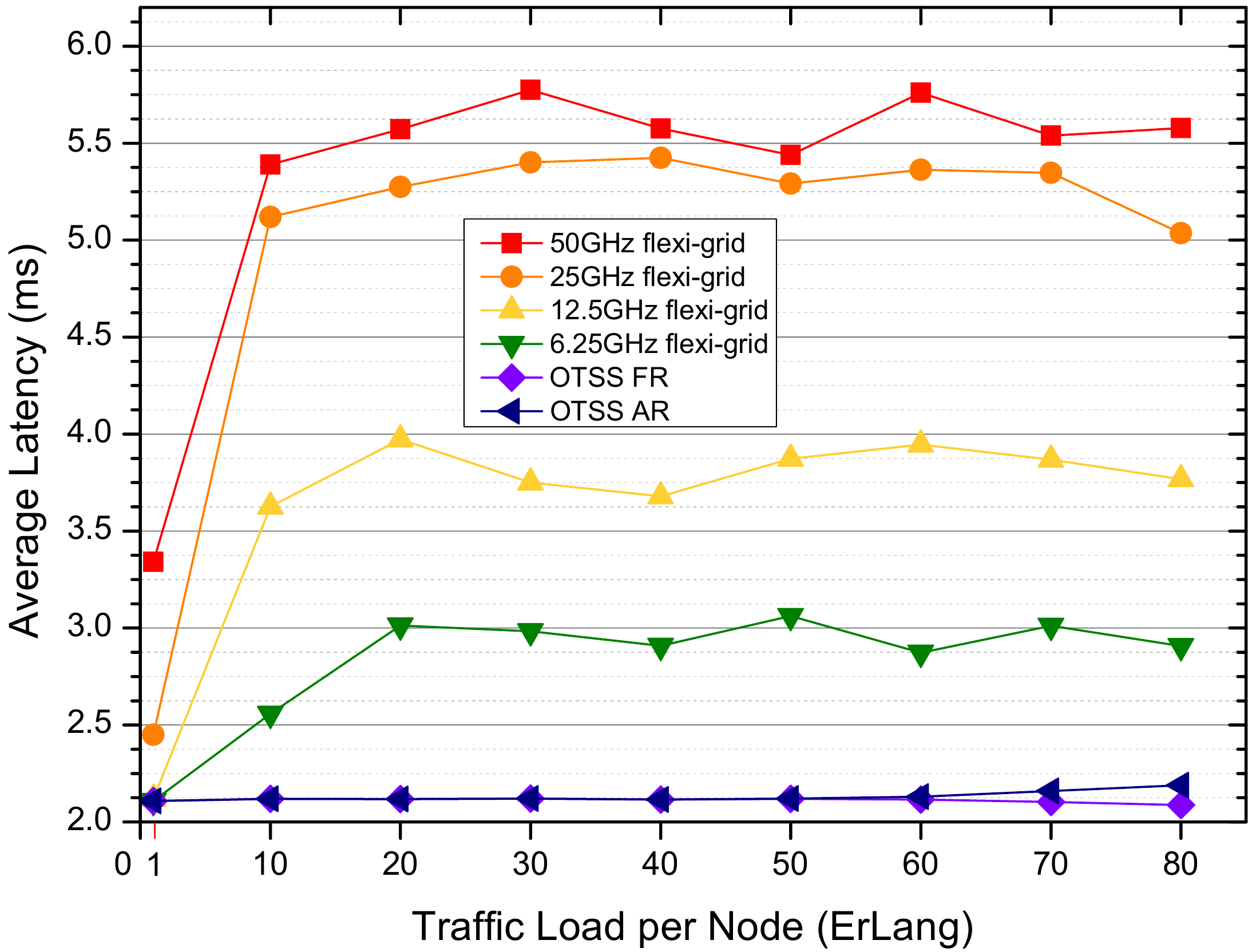}}
\hspace{0.1em}
\subfloat[Blocking probability vs. traffic load.]{
\includegraphics[width=2.25in]{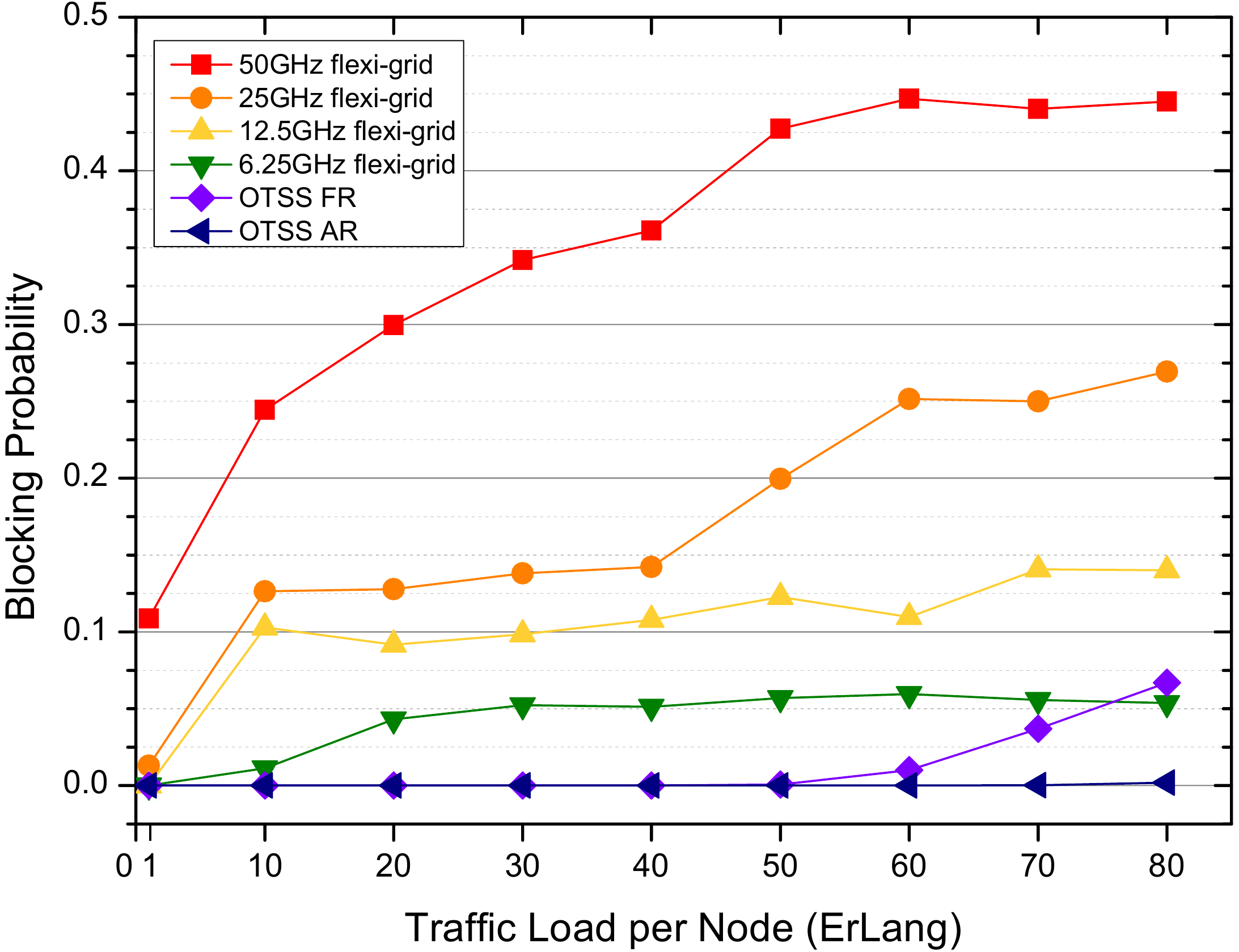}}
\caption{Dynamic performance for reliability-related traffic within 50GHz reserved spectrum.}
\label{fig:subfig} 
\end{figure*}

To verify the effectiveness of the proposed OTSS-assisted architecture, an event-driven dynamic network simulator based on C++ has been developed. As grooming latency is relatively dominant in our case, we adopt MinTHV grooming policy which has the smallest grooming nodes for flexi-grid routing and spectrum assignment to reduce latency\cite{zhu-jsac-2002}. For OTSS routing and time slot assignment, we adopt the shifting-time-slice algorithm with Fixed Routing (FR) introduced in \cite{otss-acp}. We further enrich the algorithm by considering Alternate Routing (AR) enabled by K-shortest path algorithm (K=5). Note that if a calculated path exceeds the upper latency limit (10 ms), this path cannot be used for traffic accommodation, and may result in traffic blockings.

Fig. 3(a) depicts the relationship between maximum latency and traffic load. We can find that conventional flexi-grid networks provide just-right latency under 10 ms restrictions, while OTSS can provide connections with maximum latency no longer than 4 ms under light load, and 6.4 ms under heavy load. Fig. 3(b) shows the how average latency performs as traffic load increases. We can conclude from the figure that OTSS can provide ultra low average latency (around 2.1 ms) as traffic load increases, while flexi-grid networks already suffer from high average latency. In Fig. 3(c), we further show that OTSS can endure heavier traffic burden than conventional flexi-grid networks. In most cases, even the 6.25GHz slot operates with about 6\% blocking, while OTSS with Alternate Routing (AR) achieves zero blocking. Note that, as traffic bandwidth is very small compared with lightpath capacity, blocking is mainly caused by the latency upper limit.

In conclusion, all three figures support that OTSS can provide low latency as well as low blocking probability when conventional flexi-grid networks fail to work under both light load (10 ErLangs) and heavy toad (80 ErLangs) conditions. This is achieved by the fine resource granularity of OTSS.

\section{Conclusion and Remarks}
In this paper, we proposed an OTSS-assisted optical network architecture for latency-sensitive smart-grid communications. Latency advantages of the novel OTSS networks over conventional flexi-grid optical networks are investigated. For future work, more analysis on OTSS performance in presence of both reliability-related traffic and other background loose-latency traffic will be presented.

This research opens a new area for low-latency optical networking problems. In OTSS, time dimension is exploited to better multiplex multiple fine-grained traffic requests, and precise-synchronized resource allocation scheme scheduled in time domain can guarantee that no collisions will happen, unlike Optical Burst Switching (OBS) or Optical Packet Switching (OPS). It is important to highlight that our OTSS-assisted architecture is suitable not only for smart grid, but also for other massive-connected low-latency scenarios in the future, like 5G backhaul access, ultra dense networks, Internet of Things (IoT) and intra-datacenter communications, etc. Future investigations on this important topic should be reinforced.


\section*{Acknowledgment}
This work was supported in part by Science and Technology Project of State Grid Corporation of China: The Key Technology Research of Elastic Optical Network (Grant No. 526800160006).



%

\end{document}